\documentclass{article}

\usepackage[final]{neurips_2024}

\usepackage[utf8]{inputenc} 
\usepackage[T1]{fontenc}    
\usepackage{hyperref}       
\usepackage{url}            
\usepackage{booktabs}       
\usepackage{amsfonts}       
\usepackage{nicefrac}       
\usepackage{microtype}      
\usepackage[table]{xcolor}
\usepackage{xcolor}         

\usepackage{lipsum}  
\usepackage{graphicx}
\usepackage{multirow}
\usepackage{amsmath}

\usepackage{tikz}
\newcommand*\circled[1]{\tikz[baseline=(char.base)]{
            \node[shape=circle,draw,inner sep=0.3pt, fill=c1] (char) {\textcolor{white}{#1}};}}

\definecolor{myblue}{HTML}{7BB2DD} 
\definecolor{mygray}{HTML}{DBE2E9} 
\definecolor{c1}{HTML}{0078b0} 
\definecolor{c2}{HTML}{0078b0}
\hypersetup{
    colorlinks=true,
    linkcolor=c2,
    urlcolor=c1,
    citecolor=c1,
}

\newcommand{\head}[1]{\vspace{1.7mm}\noindent{{\bf #1.}}}

\newcommand{\model}{\textsc{SynapsNet}}

\newcommand{\Se}[0]{\mathbb{S}}
\newcommand{\N}[0]{\mathcal{N}}

\title{SynapsNet: Enhancing Neuronal Population Dynamics Modeling via Learning Functional Connectivity}

\author{%
  Parsa Delavari \\
  Neuroscience\\
  University of British Columbia\\
  \texttt{parsadlr@student.ubc.ca} \\
  \And
  Ipek Oruc \\
  Ophthalmology and Visual Sciences \\
  University of British Columbia\\
  \texttt{ipor@mail.ubc.ca} \\
  \AND
  Timothy H Murphy \\
  Psychiatry \\
  University of British Columbia\\
  \texttt{thmurphy@mail.ubc.ca} \\
}

\begin{document}

\maketitle

\begin{abstract}
    The availability of large-scale neuronal population datasets necessitates new methods to model population dynamics and extract interpretable, scientifically translatable insights. Existing deep learning methods often overlook the biological mechanisms underlying population activity and thus exhibit suboptimal performance with neuronal data and provide little to no interpretable information about neurons and their interactions. In response, we introduce \model{}, a novel deep-learning framework that effectively models population dynamics and functional interactions between neurons. Within this biologically realistic framework, each neuron, characterized by a latent embedding, sends and receives currents through directed connections. A shared decoder uses the input current, previous neuronal activity, neuron embedding, and behavioral data to predict the population activity in the next time step. Unlike common sequential models that treat population activity as a multichannel time series, \model{} applies its decoder to each neuron (channel) individually, with the learnable functional connectivity serving as the sole pathway for information flow between neurons. Our experiments, conducted on mouse cortical activity from publicly available datasets and recorded using the two most common population recording modalities—Ca imaging and Neuropixels—across three distinct tasks, demonstrate that \model{} consistently outperforms existing models in forecasting population activity. Additionally, our experiments on both real and synthetic data showed that \model{} accurately learns functional connectivity that reveals predictive interactions between neurons.
    
\end{abstract}

\section{Introduction}\label{sec:intro}

Recent advancements in brain recording techniques have enabled simultaneous in-vivo recordings from hundreds of neurons. This availability of neuronal population activity has motivated numerous studies on population dynamics, which can address neuroscientific questions such as explaining brain function and behavior \citep{vyas2020computation, gallego2020long}, as well as brain decoding applications like brain-computer interfaces (BCIs) \citep{willett2023high, heelan2019decoding}. Various approaches for studying neural dynamics have been proposed in the literature, including neural manifolds \citep{duncker2021dynamics} and latent variable models (LVMs) \citep{pandarinath2018inferring, nolan2022multi}, which aim to capture low-dimensional patterns in neural activity by reproducing this activity. However, recent evidence suggests that neural activity forecasting offers great potential for studying population dynamics and capturing causal interactions in neural activity, making forecasting a promising approach for studying neural dynamics \citep{li2024amag}.

Although analyzing neuronal population dynamics holds immense potential for neuroscientific research and applications, the data is typically large, acquired from multiple animals across numerous recording sessions, necessitating novel and scalable methods for analysis and interpretation \citep{paninski2018neural, hurwitz2021building}. Deep learning models have been extensively applied to neural data to address this need \citep{richards2019deep, paninski2018neural}. However, these methods often face significant limitations, particularly in neuroscience applications. Firstly, they typically offer minimal interpretability, providing little insight into the brain's underlying mechanisms. Ideally, explainable deep learning models could harness computational power to decode vast amounts of data, with their interpretability guiding scientists toward scientific discovery, as demonstrated in successful cases \citep{lemos2023rediscovering, delavari2023artificial}. Secondly, many deep learning approaches neglect the underlying biological mechanisms \citep{kietzmann2017deep}, limiting their performance by applying generic architectures adapted from other domains to neural data. These limitations highlight the need for more biologically informed and interpretable models of neuronal population dynamics.

One potential biological prior that can be incorporated in the models' architecture is functional connectivity (FC). Functional connectivity refers to the statistical dependencies between distinct neurons/regions in the brain, indicating how they communicate and coordinate with each other during various cognitive tasks and resting states \citep{friston2011functional}. FC analysis has been utilized in a wide range of studies, including those on brain diseases and disorders \citep{du2018classification, sheline2013resting}, brain representation learning \citep{behrouz2024unsupervised}, brain classification \citep{riaz2020deepfmri, du2018classification}, and in particular neural data forecasting \citep{li2024amag}. There are various methods to infer functional connectivity across different recording modalities \citep{marzetti2019brain, yu2020benchmarking, li2009review} including spiking data \citep{perich2020inferring, ullo2014functional, pastore2018identification, quinn2011estimating}, however, each of the existing methods for inferring FC has a subset of the following limitations. They often measure only simple statistical similarities between source and target units, overlook causal interactions that are not necessarily symmetric, lack generalizability to unseen population activity, and are not flexible enough to account for multi-synaptic interactions occurring over varying delays. These challenges underscore the need for advanced methods that can more accurately capture the complex interactions within neural populations.

In this work, we introduce \model{}, a biologically inspired deep learning framework designed to model neuronal populations while uncovering functional connectivity. It integrates past neural activity, input currents from connected neurons, intrinsic neuron properties, and behavioral states to predict neural dynamics. This model builds upon previous work \citep{mi2024learning} to implicitly model dynamics of individual neurons and learns functional connectivity to capture directed interactions between neurons. We validate \model{} through extensive experiments on both synthetic and real datasets of mouse cortical activity recorded via the most common population recording methods, calcium imaging and Neuropixels \citep{jun2017fully}. Our experiments demonstrate that \model{} outperforms conventional time-series models in predicting neural activity showcasing its high capability of capturing population dynamics. Also, \model{} provides insights into the underlying interactions between neurons by accurately inferring functional connectivity.


\section{\model{}}\label{sec:synapsnet}

\subsection{Overview}

The aim of \model{} is to develop an interpretable model of neuronal populations capable of predicting future dynamics based on the current and past states of the brain and behavior. To this end, our model employs a biologically inspired framework in which a dynamical model predicts the future activity of each neuron 
by taking: \circled{1} its past activity, \circled{2} the input currents coming from the connected neurons, \circled{3} the intrinsic properties of the target neuron, and \circled{4} the animal's behavioral state (e.g., running speed). We use these components as previous research confirm their role on neuronal responses. The Hodgkin-Huxley equations \citep{hodgkin1952currents}, a foundational work in computational neuroscience, describe a neuron's membrane potential dynamics as a function of its current voltage, input currents from presynaptic neurons, and a set of neuron-specific physiological properties. Additionally, behavior has been shown to significantly impact neural responses, with the first principal components of single-trial population activity corresponding to behavioral variables such as running speed and pupil size in various brain regions, including the visual cortex \citep{stringer2019spontaneous}.

\begin{figure}[ht]
  \centering
  \resizebox{0.9\textwidth}{!}{
  \includegraphics[]{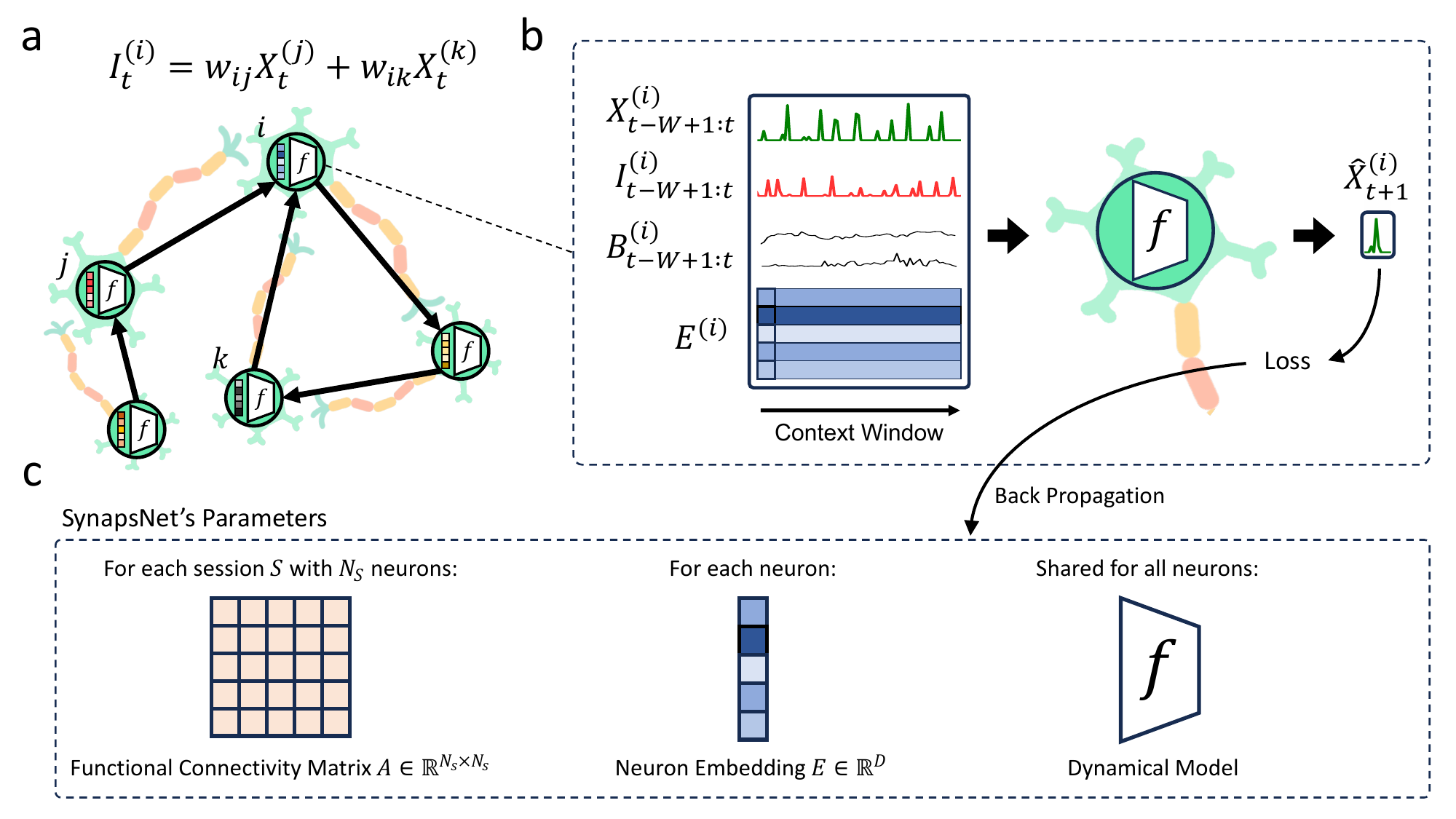}}
  \caption{\model{} overview. (a) The functional connectivity defined between neurons on the model and how input current is inferred based on functional connectivity and population activity (b) An example input frame to the dynamical model which includes past activity over the context window, past input current, past behavioral data, and the unique embedding of the target neuron. (c) The three sets of parameters in \model{}: adjacency matrix $A$ for each session, embedding vector $E$ for each neuron, and dynamical model.} 
  \label{fig:synapsnet}
\end{figure}

Although the Hodgkin-Huxley point-neuron model can accurately predict neuron dynamics, it requires detailed information on the activity of all presynaptic neurons, their synaptic strengths, and an exhaustive physiological description of the neurons. This level of detail is nearly impossible to obtain with current population recording techniques, which typically capture only a limited subset of neurons and provide minimal to no information on the population's physiological and anatomical properties. In this context, functional connectivity has been used as an alternative to anatomical connectivity for modeling interactions between neurons \citep{poli2015functional}. These functional interactions, which can be inferred from population activity, have been shown to predict neural responses even when only a small number of neurons are recorded \citep{stevenson2012functional}. Therefore, we estimate each neuron's input current based on the activity of other neurons and the population's functional connectivity. To address the lack of detailed physiological features of neurons, we use NeuPRINT \citep{mi2024learning} to learn time-invariant embeddings for each neuron based solely on population activity. The innovative approach proposed in NeuPRINT removes the dependency on neuron-specific characteristics from the dynamical model, placing it in the input so that the dynamical model becomes independent of intrinsic properties and can be shared across all neurons.

\subsection{Implementation}\label{sec:implementation}

We represent population datasets as a set of recording sessions ${\Se}$, where each session $S_j \in \Se$ includes a set of recorded neurons $\N_j$, population activity over time $X_j \in \mathbb{R}^{|\N_j| \times T}$, and available behavioral data over time $B_j \in \mathbb{R}^{N_B \times T}$. Let $M = |\bigcup\limits_{j=1}^{|\Se|} \N_j|$ be the total number of unique neurons across all recording sessions, acknowledging that some neurons may appear in multiple sessions. We define three sets of learnable parameters in \model{} (Figure~\ref{fig:synapsnet}c): \circled{1} neuron embeddings $E \in \mathbb{R}^{M \times D}$, where $D$ is the dimensionality of the embedding vectors, \circled{2} functional connectivity matrices ${\{A_j\}}_{j=1}^{|\Se|}$, where $A_j \in \mathbb{R}^{|\N_j| \times |\N_j|}$, and \circled{3} a dynamical model $f(.)$ shared among all neurons across all sessions. For simplicity, we drop the session number subscript $j$ in the following descriptions and provide expressions for a single session.

\head{Inferred input current calculation} Using population activity and functional connectivity, we calculate the input current to neuron $i$ at time $t$ as $I_t^{(i)} = \sum\limits_{k=1}^{|\N|} A^{(i,k)} X_t^{(k)}$ (Figure~\ref{fig:synapsnet}a).

\head{Dynamical model} We express the neuronal dynamics as
\[
\hat{X}_{t+1}^{(i)} = f(X_{t-W+1:t}^{(i)}, I_{t-W+1:t}^{(i)}, B_{t-W+1:t}, E^{(i)})
\]
where $W$ is the context window that determines the history considered by the dynamical model (Figure~\ref{fig:synapsnet}b). The dynamical model $f$ can be any sequential model designed for processing time-series data. The results presented in this paper are obtained using a GRU \citep{cho2014learning} (for results with other sequential models, see the supplementary materials).

\head{Training} We formulate an optimization problem using the loss $\mathcal{L}$, calculated based on the ground truth and predicted activity, to learn the three sets of parameters in \model{}:
\[
\min\limits_{f, \{A_j\}, E} \hspace{3pt} \left( \hspace{3pt} \mathbb{E}_X \left[\hspace{3pt}\mathcal{L}(\hat{X}_{t+1}^{(i)}, X_{t+1}^{(i)})\hspace{3pt}\right] \hspace{3pt} + \hspace{3pt} \lambda \hspace{1pt}\sum\limits_{j} \hspace{3pt} ||A_j||_2 \right)
\]
where $\mathcal{L}(.)$ is mean squared error (MSE), $\lambda$ is the regularization weight for preventing overfitting through $\{A_j\}_j$ and $||\hspace{3pt}.\hspace{3pt}||_2$ denotes L2 norm. For more details about the training procedure and the hyperparameters used, please refer to the supplementary materials.

\section{Experiments}\label{sec:exps}
\subsection{Population Activity Forecasting}

We evaluate \model{}'s performance in forecasting neuronal population activity using two distinct public datasets: \cite{bugeon2022transcriptomic} and \cite{siegle2021survey}.

\subsubsection*{Datasets and Preprocessing}

The dataset from \cite{bugeon2022transcriptomic} consists of two-photon calcium imaging data recorded from the mouse primary visual cortex (V1). Mice were exposed to three types of visual stimuli: natural scenes, drifting gratings, and blank screens (spontaneous activity). This dataset includes calcium traces from four mice, recorded across 17 sessions of approximately 20 minutes each. Each session captured between 178 and 868 neurons, totaling 9728 neurons. Neurons were recorded at a 4.3 Hz sampling rate on 7 imaging planes at different cortical depths, resulting in volumetric recordings. The dataset also provides neuron positions, cell types, and behavioral data such as running speed and pupil size. We normalized each neuron's calcium trace and behavioral variables.

The dataset from \cite{siegle2021survey} includes electrophysiology spiking data from mouse cortical neurons, recorded using six Neuropixels probes \citep{jun2017fully}. This dataset primarily covers V1 and higher visual areas, as well as deep subcortical areas like the LGN and hippocampus CA1. It features two session types: "brain observatory" sessions for natural scenes and drifting gratings, and "functional connectivity" sessions for spontaneous activity (blank screen). We included sessions where neuron locations were available, resulting in 46 sessions (46 animals) and 31,408 recorded neurons. The dataset also provides running speed data. We binned the spikes with a 33.3 ms bin width and normalized each neuron's response and behavioral data.

\subsubsection*{Benchmark Models}

We compare \model{} with the following existing models: standard sequential models (RNN \cite{elman1990finding}, GRU \cite{cho2014learning}, LSTM \cite{hochreiter1997long}); NeuPRINT \cite{mi2024learning}, a self-supervised method of learning neuronal representations based on population dynamics; LFADS\cite{pandarinath2018inferring}, an RNN based variational auto-encoder designed specifically for neuronal population dynamics; GWNet \cite{wu2019graph}, a graph neural network designed to model both spatial interactions and temporal patterns in data. Each sequential model was trained separately on every session due to varying model and input sizes based on recorded neurons. NeuPRINT and \model{} were trained on multiple sessions and animals due to their design, allowing the dynamical model to operate at the single-neuron level. For a fair comparison, we trained NeuPRINT and \model{} on both single-session data and the entire dataset, reporting results separately.

\subsubsection*{Evaluation on Population Activity Forecasting}

We assessed \model{} and the benchmark models on two different data modalities across three tasks. We measured test MSE loss and Pearson's correlation between true and predicted population activity, averaging over all times for each neuron and across all neurons in each recording session. The means and standard errors were calculated across sessions (Ca imaging: $n=17$; Neuropixels natural scenes and drifting gratings: $n=22$; Neuropixels spontaneous: $n=24$). Performance measures are summarized in Table~\ref{tab:comparison-result}.

NeuPRINT outperformed the conventional time-series models, highlighting the advantage of modeling dynamics at the single-neuron level over treating the population activity as multichannel time-series data. This may be due to the sparse and stochastic nature of spiking activity, which complicates population-level analysis. \model{} consistently surpassed other general models, including NeuPRINT, across both data modalities and all three tasks, demonstrating its superior capability in modeling neuronal population dynamics. Both NeuPRINT and \model{} demonstrated improved performance when trained on multiple sessions, indicating their scalability. \model{} achieved approximately 15\% higher correlation scores compared to the general models and 5\% higher compared to NeuPRINT on calcium imaging data, translating to around 80\% and 15\% relative improvements, respectively. For the Neuropixels modality, these values were about 8\% and 3\%, corresponding to approximately 30\% and 15\% relative improvements. Notably, \model{} exhibited greater improvements on tasks involving visual stimuli compared to spontaneous activity, highlighting the importance of stimulus-driven functional connectivity inferred by the model.

\begin{table*}[ht]
\caption{Performance on neural data forecasting. Mean correlation and loss (\%) $\pm$ standard error of the mean$^\dagger$.}
\label{tab:comparison-result}
\vspace{5pt}
\resizebox{1\textwidth}{!}{
\begin{tabular}{c @{\hskip 0.35in} l @{\hskip 0.35in} c c  c  c c  c  c c}
 \toprule
 \multirow{2}{*}{Data Modality} & \multirow{2}{*}{Model} & \multicolumn{2}{c}{Natural Scenes} &  & \multicolumn{2}{c}{Drifting Gratings} &  & \multicolumn{2}{c}{Spontaneous}\\
  \cmidrule(lr){3-4} \cmidrule(lr){6-7} \cmidrule(lr){9-10} 
  &  & $Corr (\%) \uparrow$ & $Loss \downarrow$ & & $Corr (\%) \uparrow$ & $Loss \downarrow$ & & $Corr (\%) \uparrow$ & $Loss \downarrow$ \\
 \midrule 
 \midrule
 
 \multirow{7}{*}{\rotatebox[origin=c]{0}{Ca Imaging}}    & \textsc{RNN} & $24.09_{\pm 0.88}$ & $0.920_{\pm 0.040}$ & & $19.95_{\pm 0.81}$ & $1.017_{\pm 0.050}$ & & $13.15_{\pm 0.75}$ & $0.924_{\pm 0.017}$ \\
                                                         & \textsc{GRU} & $24.21_{\pm 0.87}$ & $0.915_{\pm 0.040}$ & & $20.08_{\pm 0.89}$ & $1.009_{\pm 0.049}$ & & $13.31_{\pm 0.85}$ & $0.920_{\pm 0.017}$ \\
                                                         & \textsc{LSTM} & $24.60_{\pm 0.87}$ & $0.911_{\pm 0.040}$ & & $20.47_{\pm 0.85}$ & $1.006_{\pm 0.049}$ & & $13.97_{\pm 0.90}$ & $0.917_{\pm 0.017}$ \\
                                                         & \textsc{LFADS} & $25.70_{\pm 0.73}$ & $1.056_{\pm 0.040}$ & & $22.96_{\pm 0.62}$ & $1.138_{\pm 0.049}$ & & $12.33_{\pm 0.79}$ & $0.941_{\pm 0.017}$ \\
                                                         & \textsc{GWNet} & $33.04_{\pm 0.87}$ & $0.829_{\pm 0.040}$ & & $29.89_{\pm 0.85}$ & $0.910_{\pm 0.049}$ & & $28.20_{\pm 0.90}$ & $0.874_{\pm 0.017}$ \\
                                                         & \textsc{NeuPRINT} & \cellcolor{blue!20}$33.82_{\pm 0.85}$ & \cellcolor{blue!20}$0.884_{\pm 0.039}$ & & \cellcolor{blue!20}$31.17_{\pm 1.00}$ & \cellcolor{blue!20}$0.973_{\pm 0.052}$ & & \cellcolor{blue!20}$28.82_{\pm 1.25}$ & \cellcolor{blue!20}$0.867_{\pm 0.015}$ \\
                                                         & \textsc{SynapsNet} & \cellcolor{blue!20}$\mathbf{37.43_{\pm 1.05}}$ & \cellcolor{blue!20}$\mathbf{0.846_{\pm 0.037}}$ & & \cellcolor{blue!20}$\mathbf{36.14_{\pm 1.05}}$ & \cellcolor{blue!20}$\mathbf{0.927_{\pm 0.48}}$ & & \cellcolor{blue!20}$\mathbf{30.60_{\pm 1.29}}$ & \cellcolor{blue!20}$\mathbf{0.855_{\pm 0.016}}$ \\ 
                                                         
                                                         \cmidrule{2-10}
                                                         
                                                         & \textsc{NeuPRINT} (multi-session) & $34.16_{\pm 0.91}$ & $0.865_{\pm 0.038}$ & & $31.64_{\pm 1.38}$ & $1.045_{\pm 0.014}$ & & $29.25_{\pm 1.43}$ & $0.903_{\pm 0.014}$ \\
                                                         & \textsc{SynapsNet} (multi-session) & \cellcolor{blue!20}$\mathbf{37.94_{\pm 1.15}}$ & \cellcolor{blue!20}$\mathbf{0.825_{\pm 0.035}}$ & & \cellcolor{blue!20}$\mathbf{36.79_{\pm 1.62}}$ & \cellcolor{blue!20}$\mathbf{0.992_{\pm 0.015}}$ & & \cellcolor{blue!20}$\mathbf{31.60_{\pm 1.58}}$ & \cellcolor{blue!20}$\mathbf{0.886_{\pm 0.015}}$ \\
                                                         
                                                         \midrule
 \multirow{7}{*}{\rotatebox[origin=c]{0}{NeuroPixels}}   & \textsc{RNN} & $18.50_{\pm 0.30}$ & $0.963_{\pm 0.012}$ & & $20.20_{\pm 0.35}$ & $0.992_{\pm 0.032}$ & & $16.67_{\pm 0.44}$ & $0.960_{\pm 0.015}$ \\
                                                         & \textsc{GRU} & $18.11_{\pm 0.32}$ & $0.963_{\pm 0.012}$ & & $20.33_{\pm 0.37}$ & $0.989_{\pm 0.032}$ & & $16.68_{\pm 0.47}$ & $0.989_{\pm 0.015}$ \\
                                                         & \textsc{LSTM} & $18.02_{\pm 0.33}$ & $0.963_{\pm 0.012}$ & & $20.39_{\pm 0.39}$ & $0.987_{\pm 0.031}$ & & $16.74_{\pm 0.47}$ & $0.987_{\pm 0.015}$ \\
                                                         & \textsc{LFADS} & $18.26_{\pm 0.73}$ & $0.987_{\pm 0.040}$ & & $20.65_{\pm 0.62}$ & $0.953_{\pm 0.049}$ & & $17.38_{\pm 0.79}$ & $0.927_{\pm 0.017}$ \\
                                                         & \textsc{GWNet} & $21.10_{\pm 0.87}$ & $0.945_{\pm 0.040}$ & & $22.54_{\pm 0.85}$ & $0.935_{\pm 0.049}$ & & $20.21_{\pm 0.90}$ & $0.941_{\pm 0.017}$ \\
                                                         & \textsc{NeuPRINT} & \cellcolor{blue!20}$21.68_{\pm 0.21}$ & \cellcolor{blue!20}$0.942_{\pm 0.011}$ & & \cellcolor{blue!20}$22.93_{\pm 0.25}$ & \cellcolor{blue!20}$0.967_{\pm 0.029}$ & & \cellcolor{blue!20}$20.79_{\pm 0.48}$ & \cellcolor{blue!20}$0.967_{\pm 0.014}$ \\
                                                         & \textsc{SynapsNet} & \cellcolor{blue!20}$\mathbf{24..38_{\pm 0.28}}$ & \cellcolor{blue!20}$\mathbf{0.926_{\pm 0.011}}$ & & \cellcolor{blue!20}$\mathbf{25.55_{\pm 0.31}}$ & \cellcolor{blue!20}$\mathbf{0.953_{\pm 0.029}}$ & & \cellcolor{blue!20}$\mathbf{22.51_{\pm 0.53}}$ & \cellcolor{blue!20}$\mathbf{0.953_{\pm 0.014}}$ \\
                                                         
                                                         \cmidrule{2-10}
                                                         
                                                         & \textsc{NeuPRINT} (multi-session) & $22.10_{\pm 0.19}$ & $0.951_{\pm 0.012}$ & & $23.49_{\pm 0.31}$ & $0.941_{\pm 0.011}$ & & $21.31_{\pm 0.42}$ & $0.919_{\pm 0.014}$ \\
                                                         & \textsc{SynapsNet} (multi-session) & \cellcolor{blue!20}$\mathbf{25.48_{\pm 0.23}}$ & \cellcolor{blue!20}$\mathbf{0.932_{\pm 0.012}}$ & & \cellcolor{blue!20}$\mathbf{26.80_{\pm 0.031}}$ & \cellcolor{blue!20}$\mathbf{0.921_{\pm 0.011}}$ & & \cellcolor{blue!20}$\mathbf{23.83_{\pm 0.44}}$ & \cellcolor{blue!20}$\mathbf{0.905_{\pm 0.014}}$ \\
 \midrule
 \multicolumn{10}{l}{$^\dagger$ The bold values correspond to the best performance metric separately for multi-/single-session training and each data modality.}\\
 \multicolumn{10}{l}{Blue cells indicate significant difference with the next best-performing model ($p\text{-value} \leq 0.05$ achieved by paired-$t$-tests).}
\end{tabular}}
\end{table*}

\begin{figure}[ht]
  \centering
  \resizebox{0.8\textwidth}{!}{
  \includegraphics[]{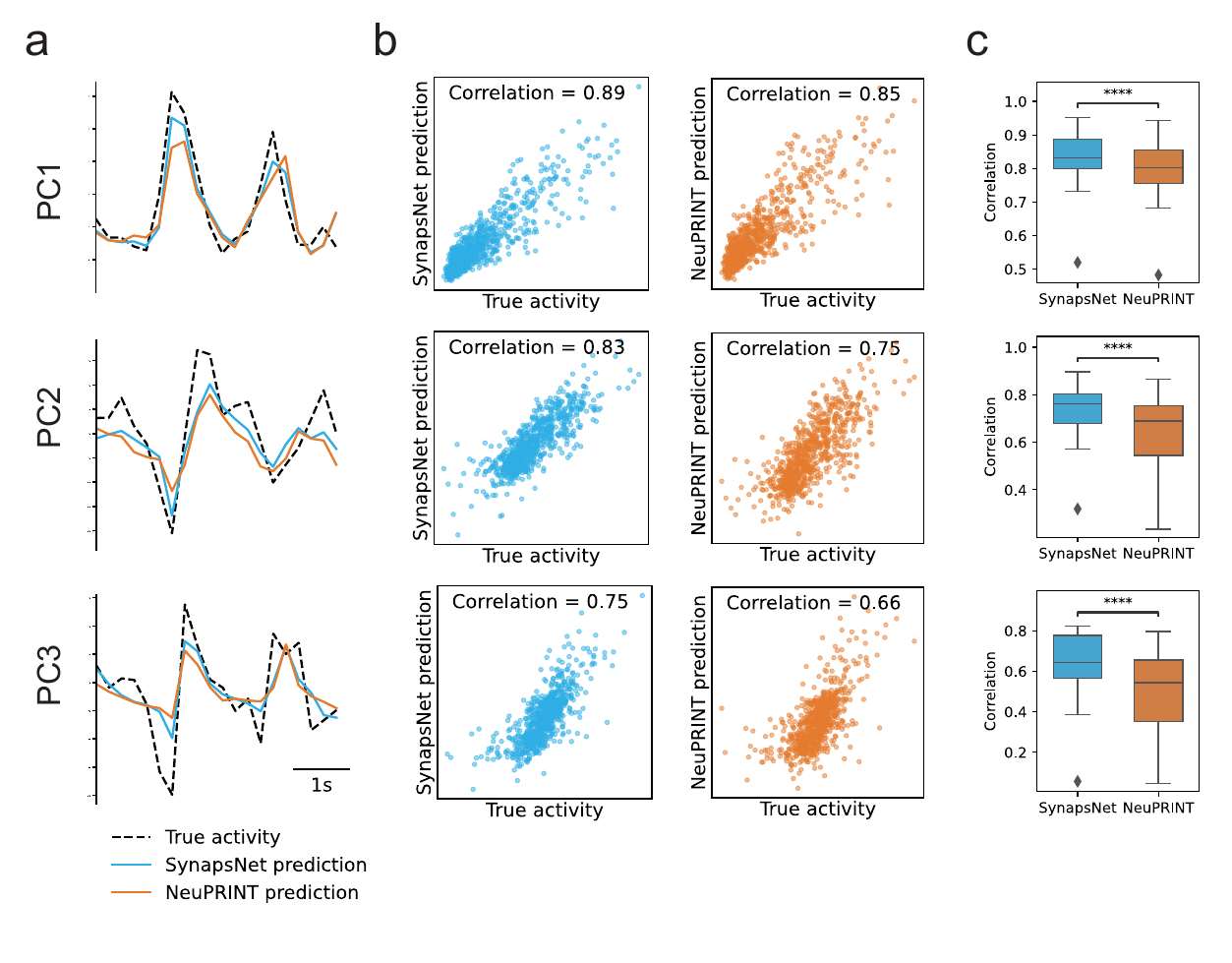}}
  \caption{Performance on neural data forecasting. (a) The first three principal components (PCs) of the true and predicted population activity sampled from an example session. (b) Correlation between the first three PCs of the true and predicted activity for the all-time points in the test set of an example session. (c) Comparison between prediction correlations achieved by SynapsNet and NeurPRINT across the test portion of all sessions. **** indicates $p\text{-value} \leq .0001$ achieved by paired-$t$-test.}
  \label{fig:pca}
\end{figure}

We also assessed the performance of \model{} at the population level. Figure~\ref{fig:pca} compares the first three principal components (PCs) of the true and predicted population activity. Across all sessions, \model{}'s predictions achieved mean correlation scores of approximately 85\%, 75\%, and 65\% for the first three PCs, respectively. These correlations were significantly higher than those achieved by NeuPRINT.

\subsubsection*{Ablation Study}
We conducted ablation experiments on \model{} by removing each of its main components. The results, presented in Table~\ref{tab:ablation}, indicate that removing functional connectivity leads to the most significant increase in test loss across both data modalities. Furthermore, the model without functional connectivity exhibits the lowest correlation score on the Neuropixels dataset and nearly matches the lowest correlation score on the calcium imaging data. These results suggest the significant role of functional connectivity in capturing the population dynamics.

\begin{table*}[h]
\caption{Ablation experiments on neural data forecasting. Mean correlation and loss (\%) $\pm$ standard error of the mean$^\dagger$.}
\label{tab:ablation}
\vspace{5pt}
\centering{
\resizebox{0.85\textwidth}{!}{
\begin{tabular}{c @{\hskip 0.35in} l @{\hskip 0.35in} c c  }
 \toprule
 \multirow{2}{*}{Data Modality} & \multirow{2}{*}{Model} & \multicolumn{2}{c}{Natural Scenes} \\
  \cmidrule(lr){3-4}
  &  & $Corr (\%) \uparrow$ & $Loss \downarrow$  \\
 \midrule 
 \midrule
 
 \multirow{4}{*}{\rotatebox[origin=c]{0}{Ca Imaging}}    & \textsc{SynapsNet} & $37.94_{\pm 1.15}$ & $0.825_{\pm 0.035}$  \\
 \cmidrule{2-4}
 & Without Neuronal Embeddings & $35.96_{\pm 1.04}$ & $0.836_{\pm 0.035}$  \\
 & Without Previous Activity & $\mathbf{33.27_{\pm 1.22}}$ & $0.851_{\pm 0.036}$  \\
 & Without Functional Connectivity & $33.94.09_{\pm 0.867}$ & $\mathbf{0.920_{\pm 0.038}}$  \\

 \midrule

 \multirow{4}{*}{\rotatebox[origin=c]{0}{NeuroPixels}}   & \textsc{SynapsNet} & $25.48_{\pm 0.23}$ & $0.932_{\pm 0.012}$  \\
 \cmidrule{2-4}
 & Without Neuronal Embeddings & $24.68_{\pm 0.23}$ & $0.933_{\pm 0.009}$  \\
 & Without Previous Activity & $24.59_{\pm 0.24}$ & $0.933_{\pm 0.010}$  \\
 & Without Functional Connectivity & $\mathbf{21.85_{\pm 0.22}}$ & $\mathbf{0.949_{\pm 0.009}}$  \\
                                                         
 \midrule
  \multicolumn{4}{l}{$^\dagger$ The bold values correspond to the largest drop in the performance metric separately for}\\
  \multicolumn{4}{l}{ multi-/single-session training and each data modality.}
  
\end{tabular}}}
\end{table*}

\subsubsection*{Evaluating Functional Connectivity Learned by \model{}}

\begin{figure}[ht]
  \centering
  \resizebox{1\textwidth}{!}{
  \includegraphics[]{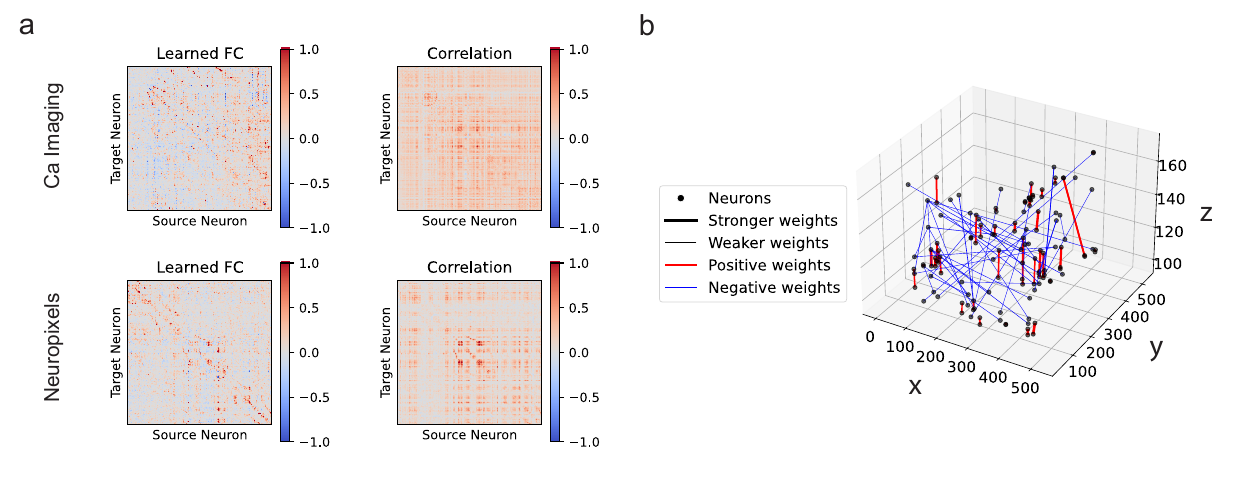}}
  \caption{Illustration of the learned functional connectivity. (a) The connectivity matrix learned by \model{} compared with that of achieved by pair-wise Pearson's correlation. The top and bottom matrices correspond to a sample session from Ca imaging and Neuropixels modalities respectively. (b) 3D visualization of the learned functional connectivity by \model{} on a sample Ca imaging session. Dots represent neurons and lines show the type and strength of the connections. Coordinates are in $\mu m$.}
  \label{fig:3d}
\end{figure}

 Figure~\ref{fig:3d}a presents the functional connectivity (FC) matrices inferred by \model{} alongside the pair-wise correlations between neurons in two example sessions—one from calcium imaging and the other from the Neuropixels dataset. \model{}'s inferred FC matrices appear sparser, more asymmetric, and more structured compared to the correlation matrices. Figure~\ref{fig:3d}b offers a 3D visualization of neurons and the learned connections, revealing clear patterns of connection types and strengths based on neuron locations, even though neuron positions were not provided to the model during training.

\begin{figure}[ht]
  \centering
  \resizebox{0.95\textwidth}{!}{
  \includegraphics[]{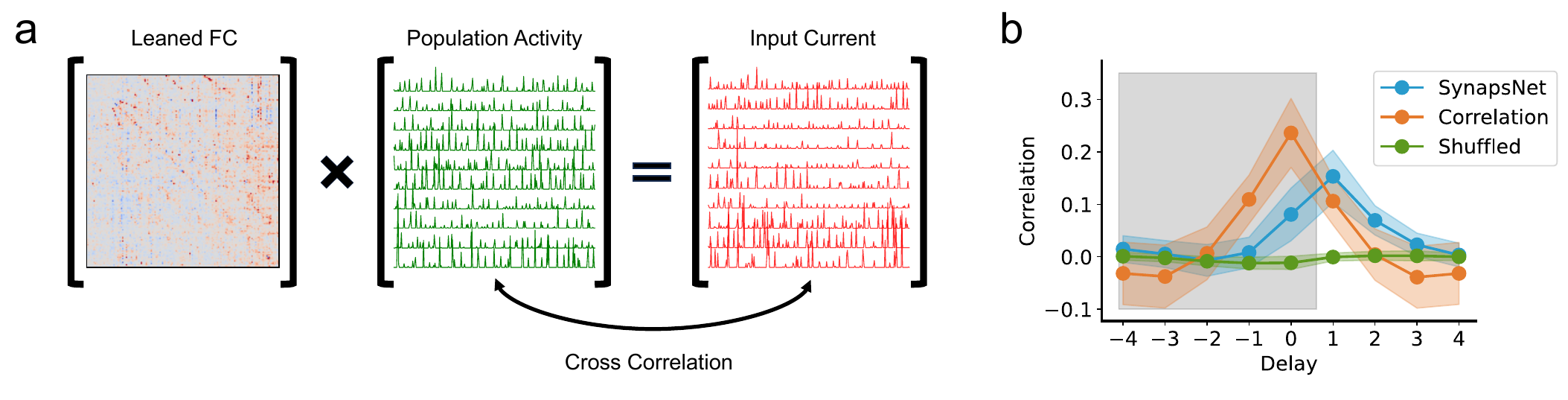}}
  \caption{Input currents achieved by functional connectivity. (a) An example calculation of the input current to each neuron based on the population activity and the learned functional connectivity. (b) Average cross-correlations between each neuron's input currents and their activity. The input currents are calculated based on connectivity matrices achieved by three different methods: \model{}, pair-wise correlation, and shuffling the \model{}'s connectivity matrix. The dots and color-shaded areas represent the mean and standard deviation of the correlations. The x-axis shows delays in time steps and the gray-shaded area marks the unavailable time points during the neuron activity forecasting task.}
  \label{fig:input_current}
\end{figure}

Additionally, we analyzed the input currents inferred by \model{}, which are latent variables used to predict future neural activity. As described in \S\ref{sec:implementation}, we calculated input currents based on population activity and functional connectivity within a recording session using the equation $I=AX$ (Figure~\ref{fig:input_current}a). We then measured the cross-correlation (correlation as a function of relative delay) between the input current to each neuron and its activity. We employed three different methods to infer functional connectivity: \model{}, pair-wise correlation, and shuffled \model{}'s FC. As illustrated in Figure~\ref{fig:input_current}b, the cross-correlation curve peaks at a positive delay, indicating that the input current is most correlated with future activity, demonstrating a predictive relationship. In contrast, the FC derived from pair-wise correlation peaks at a delay of zero. Notably, in the forecasting task, predicting activity at time $t$ relies solely on information up to time $t-1$, making current and future time points inaccessible. Interestingly, the input currents derived by \model{} show a higher correlation with population activity at positive delays compared to the pair-wise correlation method, which are the only time points available in forecasting.

Despite these promising results, evaluating learned functional connectivity remains challenging without access to ground-truth neuron connectivity. Therefore, we conducted an experiment using synthetic data simulated with known functional connectivity to further assess \model{}'s ability to capture functional dependencies in the data.

\subsection{Evaluation of Learned Functional Connectivity on Synthetic Data}

\subsubsection*{Simulation}

To further validate \model{}'s ability to recover functional interactions, we simulated the activity of a neuron population with known ground-truth connectivity (Figure~\ref{fig:synth}a). We simulated 600 interconnected neurons with connection weights randomly initialized to create a sparse connectivity matrix $A$\footnote{For better visual comparison, the locations of non-zero elements in the connectivity matrix were based on an image from the Mandelbrot set fractal to include visual patterns in the ground-truth adjacency matrix.}. We generated four random variables, two representing behavioral data ($B$) and two representing task variables ($V$). Each neuron's firing rate was calculated as: $R^{(i)}_t = \sigma \left( \Bar{R}^{(i)} + w_1\sum\limits_{k} A^{(i,k)} X^{(k)}_{t-1} + w_2B_{t-1} + w_3V_{t-1} - w_4 X^{(i)}_{t-1} + n_t \right)$ where $\sigma(.)$ is the nonlinearity ($tanh$ here), $\Bar{R}$ is the neuron's mean firing rate, $n_t$ is noise. Parameters $w_1$ to $w_4$ control the effect of input current, behavior, task, and self-inhibition respectively, and are randomly initialized for each neuron. Then, the activity of each neuron is determined by a Poisson process: $X_t^{(i)} \sim Poisson(R^{(i)}_t)$. Finally, a subset of 200 neurons is randomly chosen as "recorded" neurons. Note that the model has access to the behavioral data but not the task variables. Similar to real data, population activity and behavioral variables are normalized.

\subsubsection*{Benchmark Methods for Functional Connectivity}

We compared \model{} with two benchmark methods for inferring functional connectivity: pair-wise correlation and CURBD \citep{perich2020inferring}. Pair-wise correlation is a popular statistical method for calculating functional connectivity in neuroscience across various recording modalities. We used variants of this method with delays ($D=0,1,2$) to capture causal interactions and achieve directed FCs. CURBD is an RNN-based model designed to infer inter-region currents by learning functional connectivity in a network of interconnected neurons capable of reproducing real data.

\subsubsection*{Performance on Recovering Ground-Truth Functional Connectivity}

\begin{figure}[ht]
  \centering
  \resizebox{1\textwidth}{!}{
  \includegraphics[]{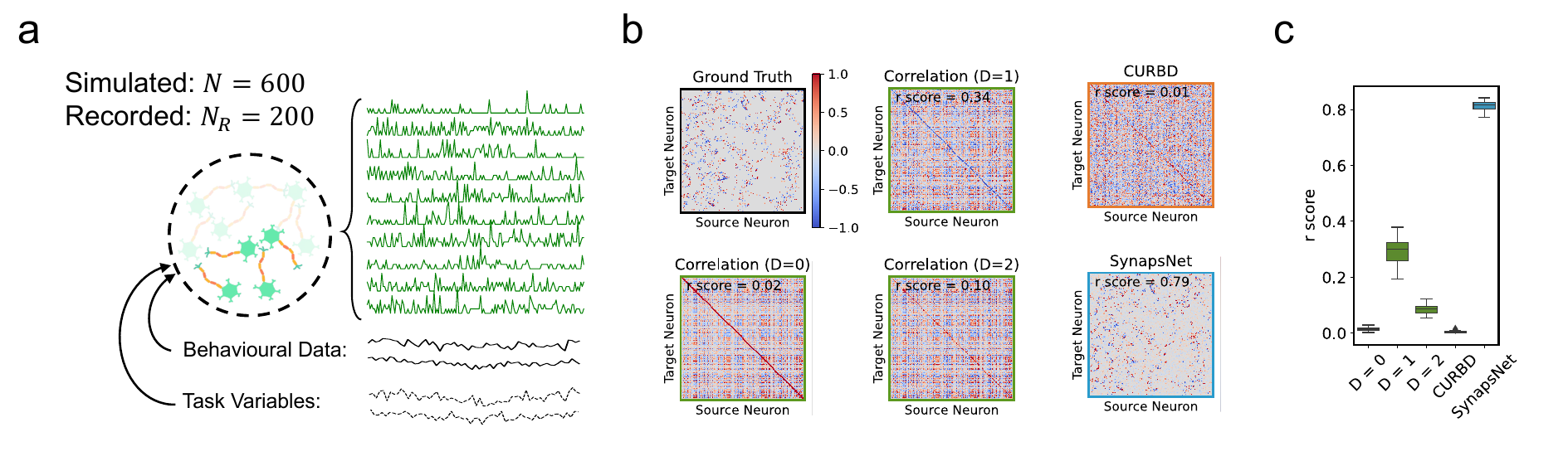}}
  \caption{Synthetic data experiment. (a) Simulation process. (b) Functional connectivity matrices inferred using \model{} and other baselines compared with the ground-truth in a single simulation run. (c) Functional connectivity reconstruction accuracies achieved by each method over 50 runs of simulation with random initializations.}
  \label{fig:synth}
\end{figure}

Figure~\ref{fig:synth}b shows the ground-truth FC alongside the inferred FC by each method in a single simulation run, while Figure~\ref{fig:synth}c compares the mean accuracy over 50 runs. \model{} reconstructed a sparse and asymmetric FC with over 80\% accuracy. CURBD, despite being trained to nearly perfectly reproduce the data (explained variance = 0.93), failed to estimate the ground-truth FC. This likely results from CURBD's RNN overfitting to the data without generalizing the learned FC to a held-out test set. Pair-wise correlation with $D=1$ partially recovered the true FC, achieving a mean accuracy of 30\%.

Notably, \model{} successfully inferred the ground-truth functional connectivity from synthetic data while its performance in predicting population activity was not perfect, with a correlation score of 36\% and a test loss of 0.83, matching its performance on real data. This suggests that \model{}'s strength in inferring FC is not constrained by the challenges of predicting highly stochastic spiking activity.

\section{Limitations}

\model{} is specifically designed to model neuronal populations, which may limit its applicability to other types of neural data. Additional research is necessary to assess its generalizability to modalities where each channel does not correspond to an individual neuron, such as electrocorticography (ECoG). Furthermore, since \model{} encodes the activity of each neuron separately, architectural modifications are required for tasks other than population activity forecasting, such as neural decoding. 

Although our experiments demonstrated promising results in analyzing the functional connectivity and input currents inferred by \model{} on both real and synthetic data, fully validating the learned connections remains challenging due to the lack of ground-truth anatomical connections in population activity datasets. Therefore, the anatomical translation of \model{}'s functional connectivity requires future validation.

\section{Conclusions}\label{sec:conclusion}

In conclusion, we introduced \model{}, a biologically inspired deep learning framework that advances neuronal population modeling by combining high predictive performance with interpretability. Our approach addresses the limitations of existing deep learning methods, which often fail to account for the biological mechanisms underlying neuronal activity and offer limited interpretability. Through extensive experiments on publicly available datasets of mouse cortical activity, recorded via calcium imaging and Neuropixels across various tasks, we showed that \model{} consistently surpassed conventional time-series models in modeling population dynamics. Additionally, our experiments on synthetic data demonstrated \model{}'s capability in identifying underlying interactions between neurons.

\section*{Supplementary Materials}

Additional details, the code used to generate the presented results, and the raw data for figures and papers are available at: \url{https://github.com/SynapsNet/SynapsNet}.



\bibliographystyle{bib_style.bst}
\bibliography{bibliography}


\newpage
\appendix
\setcounter{secnumdepth}{4}
\setcounter{tocdepth}{4}

\section{Methods Details}

\subsection{Training Details and Hyperparameters}
For each session of data, we split the population activity into three continuous partitions: approximately $90\%$ for training, $10\%$ for validation, and $10\%$ for testing. The validation set was used to select the epoch with the best performance and to tune hyperparameters, while the test set was reserved for the final evaluation of the models and reporting performance metrics.

For SynapsNet, we used context window size $W=5$, neuron embedding size of 32, and a single-layer GRU with hidden layer size of 40 as the dynamical model. During training, the training set is divided into identical (but overlapping) partitions of size $W+1$, where the first $W$ time points are the input and the last time point is the target. For batch sampling, we first select a random session from which all batch samples are chosen. This approach ensures that all samples in a batch have identical dimensionality, simplifying implementation, as different sessions contain different numbers of neurons.

For all datasets, we use a batch size of 32 and an initial learning rate of $10^{-3}$. The models are trained for 100 epochs, with the learning rate halved every 10 epochs. Additionally, dropout layers with a rate of 0.1 are employed in all neural networks. \model{} is implemented using \texttt{PyTorch} in \texttt{Python}, and evaluations are run on a Linux machine with a GPU and 16GB of RAM. Training \model{} on each session of data (single-session training) takes approximately half an hour, depending on the number of recorded neurons and the session length.

For the sake of fair comparison, we use the same training, validation, and testing data for all benchmark models and ablated versions of \model{}.

\subsection{Models Used for Comparison}

The RNN, GRU, and LSTM models each consist of 2 layers with a hidden layer size of 100. The LFADS model we used has the dimensionality of the generator, encoder, and controller all set to 256, and the dimensionality of the factor and inferred inputs to the generator set to 128. The GWNet model is defined with 32 residual channels, 32 dilation channels, 128 skip channels, and 256 end channels, with a single layer, 4 blocks, and a kernel size of 2.

\section{Additional Results}

\subsection{Sensitivity Analysis}

\begin{figure}[ht]
\centering
\begin{tabular}{ll}
\includegraphics[width=2in]{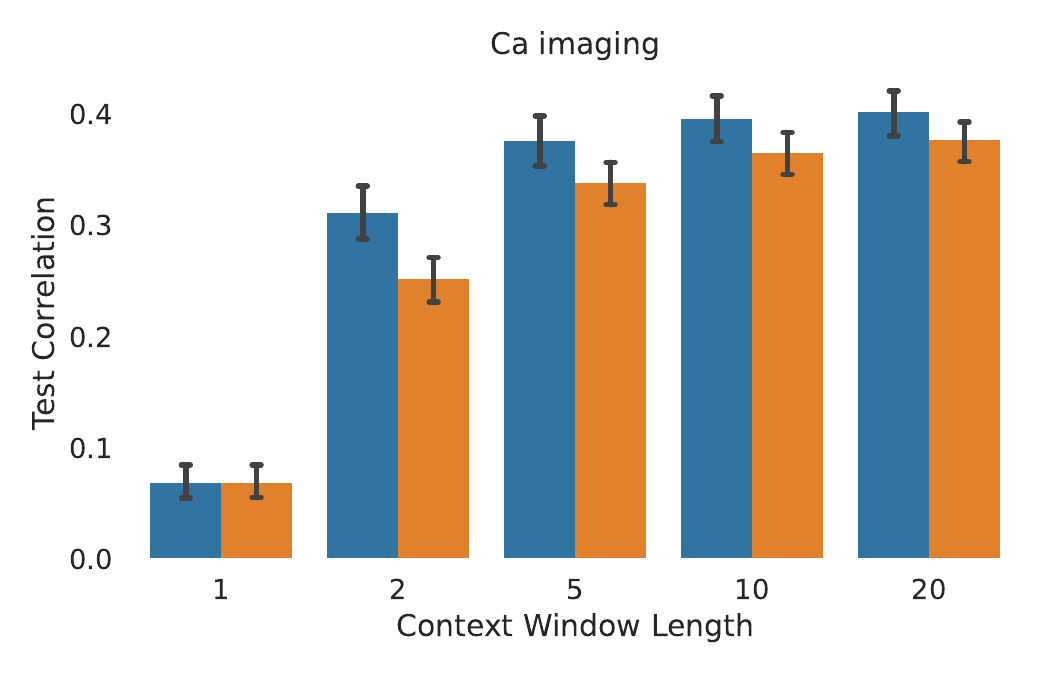} &
\includegraphics[width=2.6in]{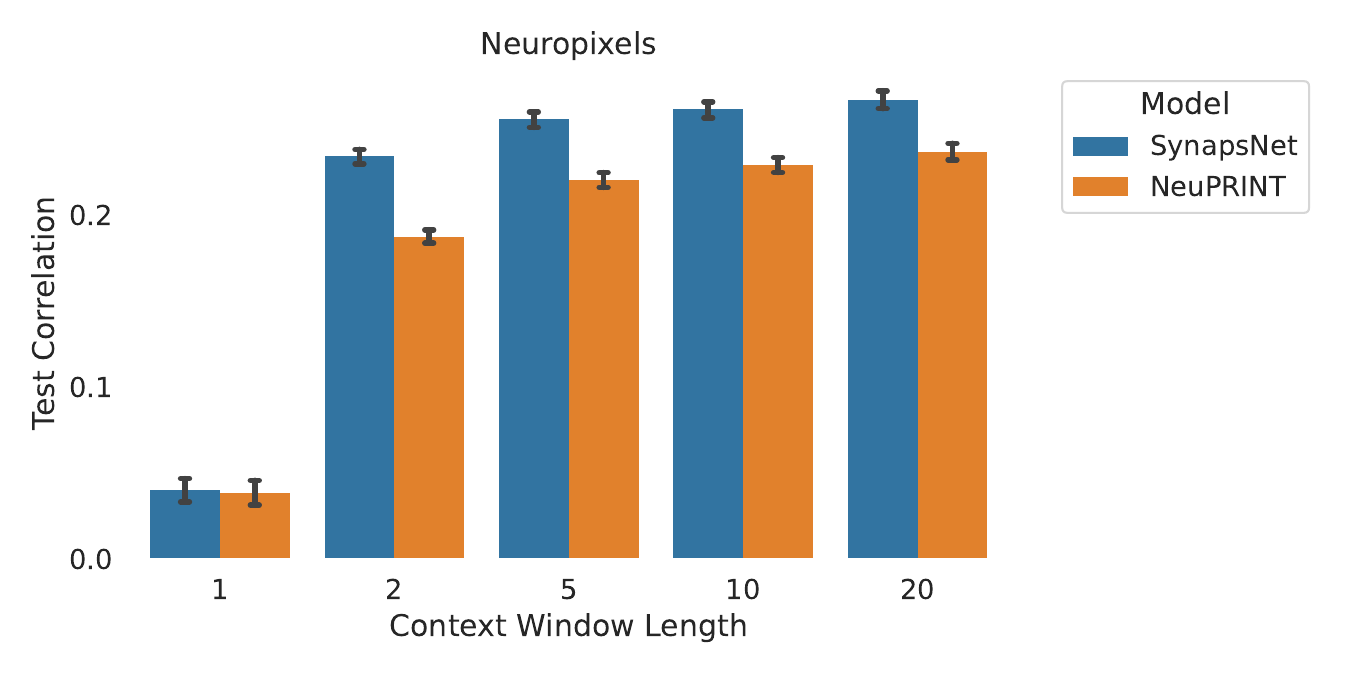} \\
\end{tabular}
\caption{Sensitivity Analysis of Context Window Length ($W$). Test correlation scores for different context window lengths are plotted separately for the Ca imaging dataset (a) and the Neuropixels dataset (b).}\label{fig:sensitivity_w}
\end{figure}

\head{Effect of Context Window Length on Forecasting Performance}
We evaluated the effect of different context window lengths ($W$) on the forecasting performance of both \model{} and NeuPRINT (the second-best-performing model). The results for Ca imaging and Neuropixels datasets are presented separately in Figure~\ref{fig:sensitivity_w}. The context window length determines how far back in the past the model can access neural activity and behavioral data to predict the population's activity at the next time point. The results indicate that at least two past time steps are required to accurately forecast neural activity, as evidenced by the substantial performance difference between $W=1$ and $W=2$. This aligns with the observations made by the authors of NeuPRINT, who identified $W=2$ as the optimal choice \citep{mi2024learning}. Performance plateaus after $W=5$, with the negligible difference observed among context window sizes of 5, 10, and 20. Therefore, all results reported in the main paper are based on $W=5$ to maintain model simplicity without significantly sacrificing performance.

\begin{figure}[ht]
\centering
\begin{tabular}{ll}
\includegraphics[width=2.in]{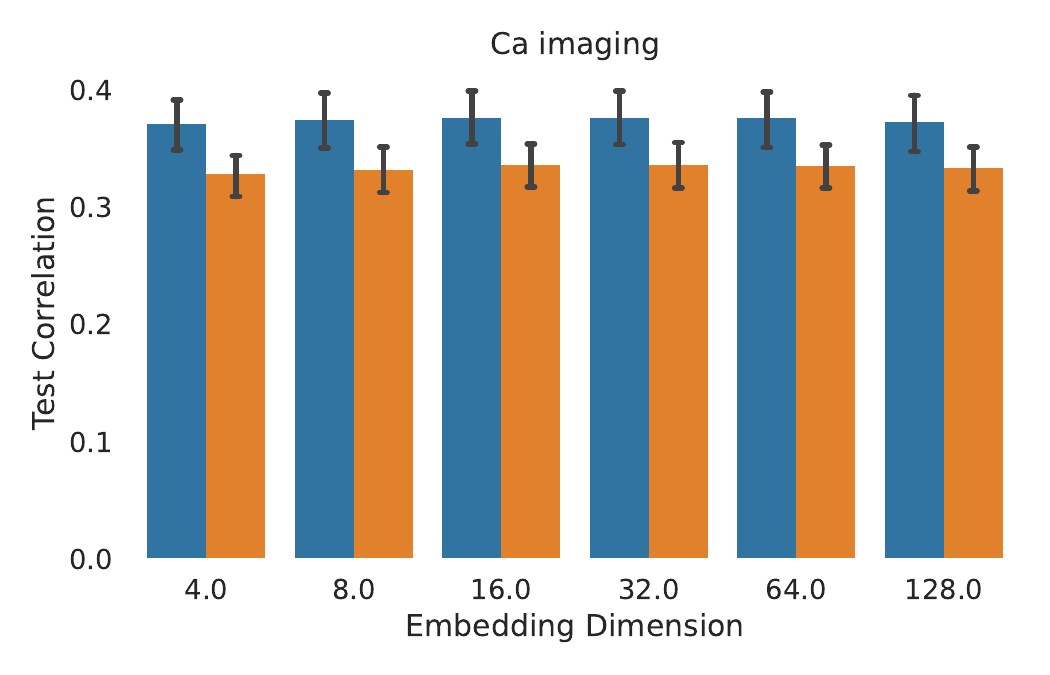} &
\includegraphics[width=2.6in]{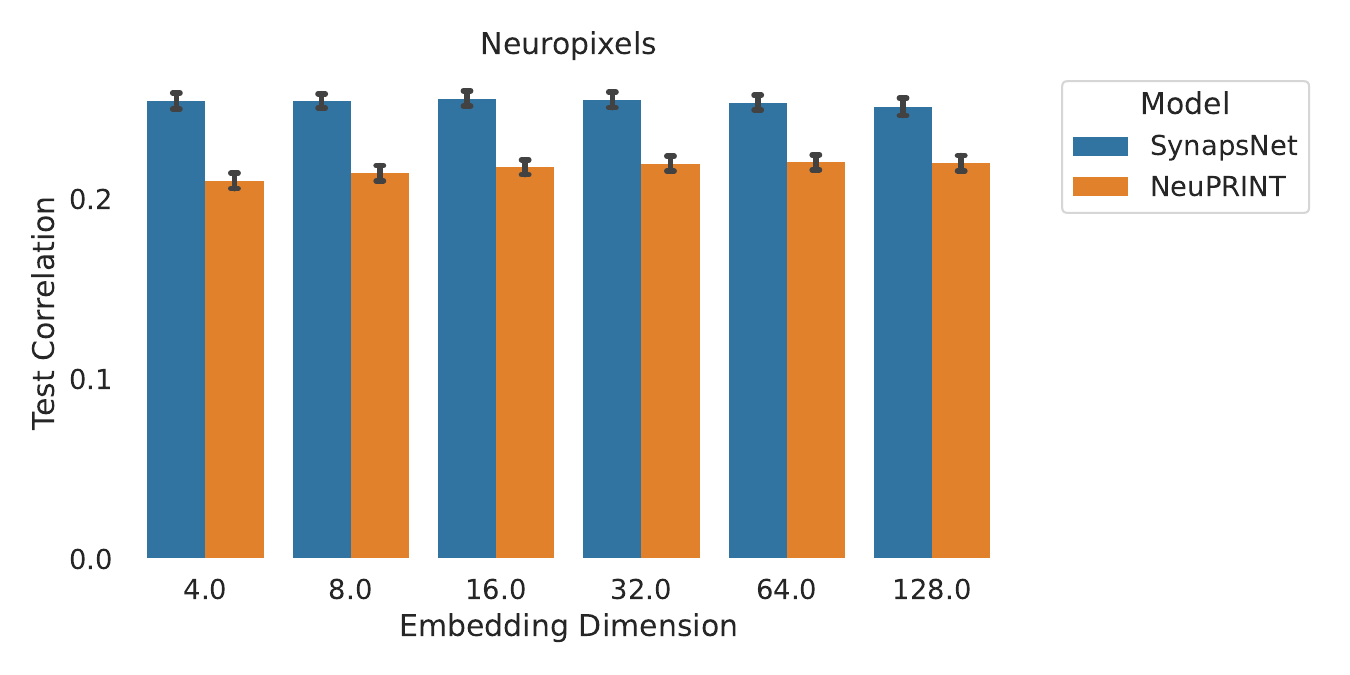} \\
\end{tabular}
\caption{Sensitivity Analysis of Neuronal Embedding Dimensionality ($D$). Test correlation scores for different neuronal embedding sizes are plotted separately for the Ca imaging dataset (a) and the Neuropixels dataset (b).}\label{fig:sensitivity_D}
\end{figure}

\head{Effect of Neuronal Embedding Dimensionality on Forecasting Performance}
We investigated the impact of varying neuronal embedding sizes ($D$) on the forecasting performance of both \model{} and NeuPRINT. Figure~\ref{fig:sensitivity_D} displays the results for the Ca imaging and Neuropixels datasets separately. Our analysis reveals that neuronal embedding dimensionality does not significantly affect neural activity forecasting accuracy. This aligns with the findings from the ablation study (Table~\ref{tab:ablation}), which indicated that neuronal embeddings contribute minimally to the performance of \model{}. Consequently, we opted for $D=32$ to remain consistent with the choice made by \cite{mi2024learning}.

\head{Effect of Using Different Sequential Models on Forecasting Performance}

We explored four different architectures for the \model{}'s dynamical model—RNN, GRU, LSTM, and Transformer—to assess their impact on the model's performance. The results for each architecture are detailed in Table~\ref{tab:sensitivity_DM} (we only used natural scenes task for simplicity). Among these, the GRU architecture achieved the highest correlation score across both data modalities, leading us to base all the results presented in this paper on the GRU dynamical model.

\begin{table*}[h]
\caption{Sensitivity Analysis of the dynamical model's architecture. Test correlation scores for different neuronal embedding sizes are reported separately for the Ca imaging dataset and the Neuropixels dataset.}
\label{tab:sensitivity_DM}
\vspace{5pt}
\centering{
\resizebox{0.85\textwidth}{!}{
\begin{tabular}{c @{\hskip 0.35in} l @{\hskip 0.35in} c c  }
 \toprule
 \multirow{2}{*}{Data Modality} & \multirow{2}{*}{Dynamical Model} & \multicolumn{2}{c}{Natural Scenes} \\
  \cmidrule(lr){3-4}
  &  & $Corr (\%) \uparrow$ & $Loss \downarrow$  \\
 \midrule 
 \midrule
 
 \multirow{4}{*}{\rotatebox[origin=c]{0}{Ca Imaging}}    & RNN & $36.99_{\pm 1.05}$ & $0.787_{\pm 0.033}$  \\
 & GRU & $37.43_{\pm 1.05}$ & $0.846_{\pm 0.037}$  \\
 & LSTM & $37.23_{\pm 1.09}$ & $0.786_{\pm 0.033}$  \\
 & Transformer (one-layer) & $29.83_{\pm 1.86}$ & $0.802_{\pm 0.036}$  \\
 & Transformer (two-layer) & $27.35_{\pm 2.48}$ & $0.826_{\pm 0.039}$  \\                
                                                         
 \midrule

 \multirow{4}{*}{\rotatebox[origin=c]{0}{Neuropixels}}    & RNN & $24.04_{\pm 0.26}$ & $0.927_{\pm 0.012}$  \\
 & GRU & $24.38_{\pm 0.28}$ & $0.926_{\pm 0.011}$  \\
 & LSTM & $23.91_{\pm 0.29}$ & $0.927_{\pm 0.012}$  \\
 & Transformer (one-layer) & $20.67_{\pm 1.32}$ & $0.917_{\pm 0.023}$  \\
 & Transformer (two-layer) & $22.00_{\pm 1.18}$ & $0.876_{\pm 0.012}$  \\         
 
 \midrule
  \multicolumn{4}{l}{$^\dagger$ The bold values correspond to the largest drop in the performance metric separately for}\\
  \multicolumn{4}{l}{ multi-/single-session training and each data modality.}
  
\end{tabular}}}
\end{table*}

\clearpage

\end{document}